\pdfoutput=1 

\documentclass[twoside,leqno,twocolumn]{article}

\usepackage[letterpaper]{geometry}

\usepackage{ltexpprt}
\usepackage{hyperref}

\usepackage{amsmath}
\usepackage{amssymb}
\usepackage[ruled,vlined,algo2e]{algorithm2e}
\usepackage[usebox=true]{jlcode}
\usepackage{bm}
\usepackage{tikz}
\usepackage[numbers]{natbib}
\usepackage{pgfplotstable}
\usepackage{xcolor}
\usepackage[normalem]{ulem}
\usepackage{multirow}
\usepackage{listings}
\pgfplotsset{compat=1.16}
\usetikzlibrary{arrows,shadows,positioning}
\definecolor{mygreen}{RGB}{80,200,120}

\tikzset{
  connection/.style={
    inner sep=0, outer sep=0,
  },
  frame/.style={
    rectangle, draw,
    text width=10em, text centered,
    minimum height=2.5em,
  },
  parallel/.style={
    rectangle, draw,
    text width=2em, text centered,
    minimum height=1.5em,drop shadow,fill=mygreen!100,
  },
  hframe/.style={
    rectangle, draw,
    text width=10em, text centered,fill=red!20,
    minimum height=2.5em,
  },
  input/.style={
    rectangle, draw,
    text width=2em, text centered,
    minimum height=1em,drop shadow,fill=red!20,
    rounded corners,
  },
  line/.style={
    draw, -latex',rounded corners=3mm,
  },
  seq/.style={
    rectangle, draw,
    minimum width=2em, text centered,
    minimum height=2em,drop shadow,fill=blue!20,
    rounded corners,
  },
  n/.style={circle, fill, minimum size=4pt,inner sep=0pt, outer sep=0pt},
}

\newcommand{\REAL}{\mathbb{R}}

\newcommand{\grad}[2]{\nabla_{#2}{#1}}

\newcommand{\hhess}[2]{\nabla^2_{#2}{#1}}

\newcommand{\cusolver}{{\tt cuSOLVER}}
\newcommand{\cusolverrf}{{\tt cuSOLVER\_RF}}

\newcommand{\KA}{{\tt KernelAbstractions.jl}}
\newcommand{\bigo}[1]{\mathcal{O}\left( #1 \right)}

\newcommand{\refalg}[1]{Algorithm~\ref{#1}}
\newcommand{\refsec}[1]{Section~\ref{#1}}
\newcommand{\reffig}[1]{Figure~\ref{#1}}

\newcommand{\reftab}[1]{Table~\ref{#1}}
\newcommand{\refeqn}[1]{(\ref{#1})}

\pgfplotstableread{data/J.dat}\matrixJ
\pgfplotstableread{data/Jc.dat}\matrixJc

\begin{document}
\newcommand\relatedversion{}


\title{\Large Batched Second-Order Adjoint Sensitivity for Reduced Space Methods\relatedversion}

\author{François Pacaud \thanks{Argonne National Laboratory}
  \and Michel Schanen \footnotemark[1]
  \and Daniel Adrian Maldonado \footnotemark[1]
\and Alexis Montoison \thanks{$GERAD$ and Polytechnique Montréal}
\and Valentin Churavy \thanks{Massachusetts Institute of Technology}
\and Julian Samaroo \footnotemark[3]
\and Mihai Anitescu \footnotemark[1]
}

\date{}
\maketitle


\fancyfoot[R]{\scriptsize{Copyright \textcopyright\ 2021 by SIAM\\
Unauthorized reproduction of this article is prohibited}}





\begin{abstract} \small\baselineskip=9pt
  This paper presents an efficient method for extracting the second-order
  sensitivities from a system of implicit nonlinear equations on upcoming graphical processing units (GPU) dominated computer systems.
  We design a custom automatic differentiation (AutoDiff) backend that
  targets highly parallel architectures by
  extracting the second-order information in batch. When the nonlinear
  equations are associated to a reduced space optimization problem, we leverage the
  parallel reverse-mode accumulation in a batched adjoint-adjoint algorithm
  to compute efficiently the reduced Hessian of the problem.
  We apply the method to extract the reduced Hessian associated to the balance equations
  of a power network, and show on the largest instances that a parallel GPU implementation is 30 times faster than a sequential CPU reference based on UMFPACK
\end{abstract}

\section{Introduction}
System of nonlinear equations are ubiquitous in numerical computing.
Solving such nonlinear systems typically depends on efficient iterative
algorithms, as for example Newton-Raphson. In this article, we are interested in the
resolution of a \emph{parametric} system of nonlinear equations, where
the solution depends on a vector of parameters $\bm{p} \in \mathbb{R}^{n_p}$.
These parametric systems are, in their abstract form, written as
\begin{equation}
  \label{eq:nonlinearsystem}
  \text{Find } \bm{x} \text{ such that } g(\bm{x}, \bm{p}) = 0  \; ,
\end{equation}
where the (smooth) nonlinear function $g: \mathbb{R}^{n_x} \times \mathbb{R}^{n_p} \to \mathbb{R}^{n_x}$
depends jointly on an unknown variable $\bm{x} \in \mathbb{R}^{n_x}$ and the parameters $\bm{p} \in \mathbb{R}^{n_p}$.

The solution $x(\bm{p})$ of \eqref{eq:nonlinearsystem} depends \emph{implicitly} on the parameters $\bm{p}$:
of particular interest are the sensitivities of the solution $x(\bm{p})$ with relation to the parameters $\bm{p}$.
Indeed, these sensitivities can be embedded inside an optimization algorithm (if $\bm{p}$ is a design variable)
or in an uncertainty quantification scheme (if $\bm{p}$ encodes an uncertainty).
It is well known that propagating the sensitivities in an iterative algorithm is nontrivial~\cite{gilbert1992automatic}.
Fortunately, there is no need to do so, as we can exploit the mathematical structure of~\eqref{eq:nonlinearsystem}
and compute directly the sensitivities of the solution~$x(\bm{p})$ using the \emph{Implicit Function
Theorem}.

By repeating this process one more step, we are able to extract second-order
sensitivities at the solution $x(\bm{p})$. However, this operation is computationally more demanding
and involves the manipulation of third-order tensors $\nabla^2_{\bm{x}\bm{x}} g, \nabla^2_{\bm{x}\bm{p}} g, \nabla^2_{\bm{p}\bm{p}} g$.
The challenge is to avoid forming explicitly such tensors by using reverse mode accumulation of second-order
information, either explicitly by using the specific structure of the problem
--- encoded by the function $g$ --- or by using automatic differentiation.

\begin{figure}
    \centering
    \begin{tikzpicture}
      \draw (-1.5, 0.7) -- (2, 0.7) -- (2, -3.5) -- (-5.0, -3.5) -- (-5.0, 0.7);
      \node at (-3.2, 0.8) {\textbf{Nonlinear system}};

      \node[frame, label=left:{$g(\bm{x}, \bm{p}) = 0$}] at (0, 0)
        {\textbf{Projection}};
      \filldraw[draw=black, fill=black] (-1, -0.7) -- (1, -0.7) -- (0, -0.9);

      \node[frame, label=left:{$\nabla_{\bm{x}}g, \nabla_{\bm{p}}g$}] at (0, -1.5)
        {\textbf{Reduced gradient}};
      \filldraw[draw=black, fill=black] (-1, -2.2) -- (1, -2.2) -- (0, -2.4);

      \node[hframe, label=left:{$\nabla^2_{\bm{x}\bm{x}}g, \nabla^2_{\bm{x}\bm{p}}g, \nabla^2_{\bm{p}\bm{p}}g$}] at (0, -3)
        {\textbf{Reduced Hessian}};

      \node (p1) at (3.5, 0) {};
     \draw[double,->] (2.1,0.0) -- node[above=.05cm,pos=.6, align=center]{$F$} (p1);

     \node (p2) at (3.5, -1.5) {};
     \draw[double,->] (2.1,-1.5) -- node[above=.05cm,pos=.5, align=center]{$\nabla_{\bm{p}} F$} (p2);

     \node (p3) at (3.5, -3.0) {};
     \draw[double,->] (2.1,-3.0) -- node[above=.05cm,pos=.5, align=center]{$\nabla^2_{\bm{p}\bm{p}} F$} (p3);
    \end{tikzpicture}
    \caption{
      Reduced space algorithm. This article focuses on the last block, in red.
      If $F$ is an objective function,
      the reduced gradient $\nabla_{\bm{p}} F$ and the reduced Hessian $\nabla^2_{\bm{p}\bm{p}} F$
      can be used in any nonlinear optimization algorithm.
    }
    \label{fig:overview}
\end{figure}
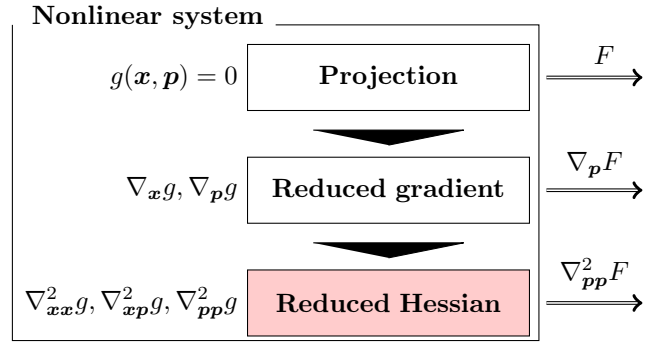
As illustrated in Figure~\ref{fig:overview},
this paper covers the efficient computation of the
second-order sensitivities of a nonlinear system~\eqref{eq:nonlinearsystem}. The sparsity structure of the problem is passed to a custom Automatic Differentiation (AutoDiff) backend that automatically generates
all the intermediate sensitivities from the implementation of $g(\bm{x},\bm{p})$. To get a tractable
algorithm, we use an adjoint model implementation of the generated
first-order sensitivities to avoid explicitly forming third-order derivative tensors.
As an application, we compute the reduced Hessian of
the nonlinear equations corresponding to the power flow balance equations of a power grid~\cite{tinney1967power}.
The problem has an unstructured graph structure, leading
to some challenge in the automatic differentiation library, that we discuss extensively.
We show that the reduced Hessian associated to the power flow
equations can be computed efficiently in parallel, by using batches of
Hessian-vector products. The underlying motivation is to embed the reduction
algorithm in a real-time tracking procedure~\cite{tang2017real}, where the reduced Hessian updates
have to be fast to track a suboptimal solution.

In summary, we aim at devising a \emph{portable}, \emph{efficient}, and easily maintainable reduced Hessian algorithm. To this end, we leverage
the expressiveness offered by the Julia programming language. Due to the algorithm's design,
the automatic differentiation backend and the reduction algorithm are transparently implemented on the GPU
without any changes to the algorithm's core implementation, thus realizing a composable software design.

\subsection{Contributions}
Our contribution is a tractable SIMD algorithm and implementation
to evaluate the reduced Hessian from a parametric system of nonlinear equations~\eqref{eq:nonlinearsystem}.
This consists of three closely intertwined components.
First, we implement the nonlinear function $g(\bm{x}, \bm{p})$
using the programming language Julia~\cite{bezanson2017julia}
and the portability layer~\KA\ to generate abstract kernels
working on various GPU architectures (CUDA, ROCm).
Second, we develop a custom AutoDiff backend on top of the portability
layer to extract automatically the first-order sensitivities $\nabla_{\bm{x}} g, \nabla_{\bm{p}} g$
and the second-order sensitivities $\nabla_{\bm{x}\bm{x}}^2 g, \nabla_{\bm{x}\bm{p}}^2g, \nabla_{\bm{p}\bm{p}}^2 g$.
Third, we combine these in an efficient parallel accumulation of the reduced
Hessian associated to a given reduced space problem.
The accumulation involves both Hessian tensor contractions and two sparse linear solves
with multiple right-hand sides.
Glued together, the three components give a generic code
able to extract the second-order derivatives from a power grid problem,
running in parallel on GPU architectures.
Numerical experiments with Volta GPUs (V100)
showcase the scalability of the approach, reaching a 30x faster computation on the largest instances when compared to a reference CPU implementation using UMFPACK. Current researches suggest that a parallel OpenMP power flow implementation using multi-threading (on the CPU alone) potentially achieves a speed-up of 3~\cite{ahmadi2018parallel} or up to 7 and 70 speed-up for Newton-Raphson and batched Newton-Raphson~\cite{2021MTNR}, respectively. However, multi-threaded implementations are not the scope of this paper as we focus on architectures where GPUs are the dominant FLOP contributors for our specific application of second-order space reduction.

\section{Prior Art}
\label{sec:priorart}
In this article we extract the second-order sensitivities from
the system of nonlinear equations using automatic differentiation (AutoDiff).
AutoDiff on Single Instruction, Multiple Data (SIMD) architectures alike the
CUDA cores on GPUs is an ongoing research effort.
Forward-mode AutoDiff effectively adds tangent components to the variables
and preserves the computational flow. In addition, a vector mode can be
applied to propagate multiple tangents or directional derivatives at once.
The technique of automatically generating derivatives of function
implementations has been investigated since the 1950s \cite{nolan1953analytical,beda1959programs}.

Reverse- or adjoint-mode AutoDiff reverses the computational flow and thus
incurs a lot of access restrictions on the final code. Every read of a
variable becomes a write, and vice versa. This leads to application-specific
solutions that exploit the structure of an underlying problem to generate
efficient adjoint code \cite{bluhdorn2020automat,grabner2008automatic,huckelheim2018parallelizable}.
Most prominently, the reverse mode is currently implemented as
backpropagation in machine learning. Indeed, the backpropagation has a long
history (e.g., \cite{bryson1962steepest})  with the reverse mode in AutoDiff
being formalized for the first time in \cite{linnainmaa1976taylor}. Because
of the  limited size and single access pattern of neural networks, current
implementations \cite{NEURIPS2019_9015,tensorflow2015-whitepaper,innes2018flux}
reach a high throughput on GPUs. For the wide field of numerical simulations,
however, efficient adjoints of GPU implementations remain challenging~\cite{10.1145/3458817.3476165}. In
this work we combine the advantages of GPU implementations of the gradient
with the evaluation of Hessian-vector products first introduced in \cite{pearlmutter1994fast}.

Reduced-space methods have been applied widely in
uncertainty quantification and partial differential equation (PDE)-constrained
optimization~\cite{biegler2003large}, and their applications in the optimization
of power grids is known since the 1960s~\cite{dommel1968optimal}.
However, extracting the second-order sensitivities in the reduced space
has been considered tedious to implement and hard to motivate on classical CPU architectures (see \cite{kardos2020reduced}
for a recent discussion about the computation of the reduced Hessian on the CPU).
To the best of our knowledge, this paper is the first to present a SIMD focused algorithm leveraging
the GPU to efficiently compute the reduced Hessian of the power flow
equations.

\section{Reduced space problem}
In \refsec{sec:background:powerflow} we briefly introduce the power flow
nonlinear equations to motivate our application. We present in \refsec{sec:background:reducedspace}
the reduced space problem associated with the power flow problem, and recall
in \refsec{sec:background:adjoint} the first-order adjoint method, used to evaluate
efficiently the gradient in the reduced space, and later applied to compute the adjoint of the sensitivities.

\subsection{Presentation of the power flow problem.}
\label{sec:background:powerflow}
We present a brief overview of the steady-state solution of the power flow
problem. The power grid can be described as a graph $\mathcal{G} = \{V, E\}$
with $n_v$ vertices and $n_e$ edges. The steady state of the network is described
by the following nonlinear equations, holding at all nodes $i \in V$,
\begin{equation}
  \label{eq:powerflow}
  \left\{
  \begin{aligned}
  P_i^{inj} &= v_i \sum_{j \in A(i)} v_j (g_{ij}\cos{(\theta_i - \theta_j)} + b_{ij}\sin{(\theta_i - \theta_j})) \,,  \\
  Q_i^{inj} &= v_i \sum_{j \in A(i)} v_j (g_{ij}\sin{(\theta_i - \theta_j)} - b_{ij}\cos{(\theta_i - \theta_j})) \,,
  \end{aligned}
  \right.
\end{equation}
where at node $i$, $(P_i^{inj}$ and  $Q_i^{inj})$ are respectively the active and
reactive power injections; $v_i$ is the voltage magnitude; $\theta_i$ the
voltage angle; and $A(i) \subset V$ is the set of adjacent nodes:
for all $j \in A(i)$, there exists a line $(i, j)$ connecting node $i$ and node $j$.
The values $g_{ij}$ and $b_{ij}$ are associated with the physical characteristics
of the line $(i, j)$. Generally, we distinguish the (\emph{PV}) nodes --- associated
to the generators --- from the  (\emph{PQ}) nodes comprising only loads.
We note that the structure of the nonlinear equations~\refeqn{eq:powerflow} depends
on the structure of the underlying graph through the adjacencies~$A(\cdot)$.

We rewrite the nonlinear equations~\refeqn{eq:powerflow}
in the standard form~\refeqn{eq:nonlinearsystem}.
At all nodes the power injection $P^{inj}_i$
should match the net production $P_i^g$ minus the load $P_i^d$:
\begin{equation}
  \label{eq:powerflowvec}
  g(\bm{x}, \bm{p}) =
    \begin{bmatrix}
      \bm{P}_{pv}^{inj} - \bm{P}^{g} + \bm{P}_{pv}^d \\
      \bm{P}_{pq}^{inj} + \bm{P}_{pq}^d \\
      \bm{Q}_{pq}^{inj} + \bm{Q}_{pd}^d
    \end{bmatrix}
    = 0
    ~ \text{with} ~
    \bm{x} =
    \begin{bmatrix}
      \bm{\theta}^{pv} \\ \bm{\theta}^{pq} \\ \bm{v}^{pq}
    \end{bmatrix} \,.
\end{equation}
In~\refeqn{eq:powerflowvec}, we have selected only a subset of the power flow equations
\refeqn{eq:powerflow} to ensure that the nonlinear system $g(\bm{x}, \bm{p}) = 0$
is invertible with respect to the state $\bm{x}$. The unknown variable $\bm{x}$
corresponds to the voltage
angles at the PV and PQ nodes and the voltage magnitudes at the PQ nodes.
However, in contrast to the variable $\bm{x}$, we have some flexibility in
choosing the parameters $\bm{p}$.

In optimal power flow (OPF) applications, we are looking at minimizing
a given operating cost $f: \mathbb{R}^{n_x} \times \mathbb{R}^{n_p} \to \mathbb{R}$
(associated to the active power generations $\bm{P}^g$) while satisfying
the power flow equations~\eqref{eq:powerflowvec}. In that particular case,
$\bm{p}$ is a design variable associated to the active power generations
and the voltage magnitude at PV nodes: ~$\bm{p} = (\bm{P}^g, \bm{v}_{pv})$.
We define the OPF problem as
\begin{equation}
  \label{eq:nonlinearopt}
  \min_{\bm{x}, \bm{p}} \; f(\bm{x}, \bm{p}) ~ \text{ subject to }
  g(\bm{x}, \bm{p}) = 0 \; .
\end{equation}

\subsection{Projection in the reduced space.}
\label{sec:background:reducedspace}
We note that in Equation~\eqref{eq:powerflowvec}, the functional $g$ is continuous
and that the dimension
of the output space is equal to the dimension of the input variable $\bm{x}$. Thanks
to the particular network structure of the problem (encoded by the adjacencies $A(\cdot)$
in \eqref{eq:powerflow}), the Jacobian $\nabla_{\bm{x}} g$ is sparse.

Generally, the nonlinear system~\eqref{eq:powerflowvec} is solved iteratively
with a Newton-Raphson algorithm.
If at a fixed parameter $\bm{p}$ the Jacobian $\nabla_{\bm{x}}g$ is invertible,
we compute the solution $x(\bm p)$ iteratively, starting from an initial guess $\bm{x}_0$:
$\bm{x}_{k+1} = \bm{x}_k - (\nabla_{\bm{x}} g_k)^{-1} g(\bm{x}_k, \bm{p})$
for $k=1,\dots, K$.
We know that if $\bm{x}_0$ is close enough to the solution, then
the convergence of the algorithm is quadratic.

With the projection completed, the optimization problem~\eqref{eq:nonlinearopt}
rewrites in the reduced space as
\begin{equation}
  \label{eq:nonlinearoptreduced}
  \min_{\bm{p}} \; F(\bm p) := f\big(x(\bm{p}), \bm{p}\big) \; ,
\end{equation}
reducing the number of optimization variables from $n_x+n_p$ to $n_p$,
while at the same time eliminating all equality constraints in the formulation.

\subsection{First-Order Adjoint Method.}
\label{sec:background:adjoint}

With the reduced space problem~\eqref{eq:nonlinearoptreduced} defined,
we compute the reduced gradient $\nabla_{\bm{p}} F$ required for the reduced space optimization routine.
By definition, as $x(\bm{p})$ satisfies $g(x(\bm{p}), \bm{p}) = 0$, the chain rule yields
$\nabla_{\bm{p}} F = \nabla_{\bm{p}} f + \nabla_{\bm{x}} f \cdot \nabla_{\bm{p}} x$
with
$\nabla_{\bm{p}} x= - \big(\nabla_{\bm{x}}g)^{-1} \nabla_{\bm{p}}g$.
However, evaluating the full sensitivity matrix $\nabla_{\bm{p}} x$ involves the
resolution of $n_x$ linear system.
On the contrary, the \emph{adjoint method} requires solving a
\emph{single} linear system.
For every dual $\bm{\lambda} \in \mathbb{R}^{n_x}$, we introduce a
Lagrangian function defined as
\begin{equation}
  \label{eq:lagrangian}
 \ell(\bm{x}, \bm{p}, \bm{\lambda}) := f(\bm{x}, \bm{p}) + \bm{\lambda}^\top g(\bm{x}, \bm{p})
 \; .
\end{equation}
If $\bm{x}$ satisfies $g(\bm{x}, \bm{p}) = 0$, then the Lagrangian $\ell(\bm{x},
\bm{p}, \bm{\lambda})$ does not depend on $\bm{\lambda}$ and we
get $\ell(\bm{x}, \bm{p}, \bm{\lambda}) = F(\bm{p})$. By
using the chain rule, the total derivative of $\ell$
with relation to the parameter $\bm{p}$ satisfies
\begin{equation*}
  \begin{aligned}
    d_p \ell &= \big(\nabla_{\bm{x}} f \cdot \nabla_{\bm{p}} x +
    \nabla_{\bm{p}} f  \big) +
    \bm{\lambda}^\top \big(\nabla_{\bm{x}} g \cdot \nabla_{\bm u} x +
    \nabla_{\bm{p}} g \big)  \\
             &= \big(\nabla_{\bm{p}} f + \bm{\lambda}^\top \nabla_{\bm{p}} g \big) +
             \big(\nabla_{\bm{x}} f + \bm{\lambda}^\top\nabla_{\bm{x}} g \big) \nabla_{\bm{p}} x  \;.
  \end{aligned}
\end{equation*}
We observe that
by setting the first-order adjoint to $\bm{\lambda} = - (\nabla_{\bm{x}}g)^{-\top}\nabla_{\bm{x}} f^\top$,
the reduced gradient $\nabla_{\bm{p}} F$ satisfies
\begin{equation}
  \label{eq:reduced_gradient}
  \nabla_{\bm{p}} F =
  \nabla_{\bm{p}} \ell =
  \nabla_{\bm{p}} f +
  \bm{\lambda}^\top \nabla_{\bm{p}} g
  \; ,
\end{equation}
with $\bm{\lambda}$ evaluated by solving a single linear system.

\section{Parallel reduction algorithm}
It remains now to compute the reduced Hessian.
We present in \refsec{sec:reduction:soc}
the adjoint-adjoint method and describe in~\refsec{sec:reduction:ad}
how to evaluate efficiently the second-order sensitivities with Autodiff. By combining together
the Autodiff and the adjoint-adjoint method, we devise in \refsec{sec:reduction:algorithm}
a parallel algorithm to compute the reduced Hessian.

\subsection{Second-Order Adjoint over Adjoint Method.}
\label{sec:reduction:soc}
Among the different Hessian reduction schemes presented in~\cite{papadimitriou2008direct} (direct-direct, adjoint-direct,
direct-adjoint, adjoint-adjoint), the \emph{adjoint-adjoint} method
has two key advantages to evaluate the reduced Hessian on the GPU.
First, it avoids forming explicitly the dense tensor $\nabla^2_{\bm{p}\bm{p}} x$
and the dense matrix $\nabla_{\bm{p}} x$, leading to important memory savings on the larger cases.
Second, it enables us to compute the reduced Hessian slice by slice, in an embarrassingly parallel
fashion.

Conceptually, the adjoint-adjoint method extends the adjoint method (see \refsec{sec:background:adjoint})
to compute the second-order derivatives $\nabla^2 f \in \mathbb{R}^{n_p\times n_p}$ of the
objective function $f((x(\bm{p}), \bm{p})$.
The adjoint-adjoint method computes the matrix $\nabla^2 f$ slice by slice,
by using $n_p$ Hessian-vector products $(\nabla^2 f) \bm{w}$ (with $\bm{w} \in \mathbb{R}^{n_p}$).

By definition of the first-order adjoint $\bm{\lambda}$,
the derivative of the Lagrangian function~\refeqn{eq:lagrangian} with respect to $\bm{x}$ is null:
\begin{equation}
  \label{eq:first_order}
    \grad{f}{\bm{x}}(\bm{x}, \bm{p}) + \bm{\lambda}^\top \grad{g}{\bm{x}}(\bm{x}, \bm{p})  = 0
    \; .
\end{equation}
Let $\hat{g}(\bm{x}, \bm{p}, \bm{\lambda}) :=  \grad{f}{\bm{x}}(\bm{x}, \bm{p}) + \bm{\lambda}^\top \grad{g}{\bm{x}}(\bm{x}, \bm{p})$.
We define a new Lagrangian associated with \refeqn{eq:first_order}
by introducing two second-order adjoints $\bm{z}, \bm{\psi} \in \mathbb{R}^{n_x}$
and a vector $\bm{w} \in \mathbb{R}^{n_p}$:
\begin{multline}
  \hat{\ell}(\bm{x}, \bm{p}, \bm{w}, \bm{\lambda}; \bm{z}, \bm{\psi}) :=
  (\nabla_{\bm{p}} \ell)^\top \bm{w} + \\ \bm{z}^\top g(\bm{x}, \bm{p})
  + \bm{\psi}^\top \hat{g}(\bm{x}, \bm{p}, \bm{\lambda})
  \; .
\end{multline}
By computing the derivative of $\hat{\ell}$ and eliminating
the terms corresponding to $\nabla_{\bm{x}} \bm{\lambda}$
and $\nabla_{\bm{p}} \bm{\lambda}$, we get the following
expressions for the second-order adjoints $(\bm{z}, \bm{\psi})$:
\begin{equation}
  \label{eq:socadjoint}
  \left\{
  \begin{aligned}
    & (\grad{g}{\bm{x}}) \bm{z} = - \big(\nabla_{\bm{p}} g\big)^\top \bm{w} \\
    & (\grad{g}{\bm{x}})^\top \bm{\psi} =
    - (\nabla^2_{\bm{x}\bm{p}} \ell) \bm{w} \;
    - (\nabla^2_{\bm{x}\bm{x}} \ell) \bm{z} \; .
  \end{aligned}
  \right.
\end{equation}
Then, the reduced-Hessian-vector product reduces to
\begin{equation}
  \label{eq:hessvecprod}
  \big(\nabla^2 f\big) \bm{w} = (\hhess{\ell}{\bm{p}\bm{p}}) \, \bm{w}
  +  (\hhess{\ell}{\bm{p}\bm{x}})^\top \bm{z}
  +  (\grad{g}{\bm{p}})^\top\bm{\psi}
  \;.
\end{equation}
As $\nabla^2 \ell = \nabla^2 f + \bm{\lambda}^\top \nabla^2 g$,
we observe that both Equations~\refeqn{eq:socadjoint}
and \refeqn{eq:hessvecprod} require evaluating the product
of the three tensors $\nabla^2_{\bm{x}\bm{x}} g,$ $\nabla^2_{\bm{x}\bm{p}} g
,$ and $\nabla^2_{\bm{p}\bm{p}} g$, on the left with the adjoint
$\bm{\lambda}$ and on the right with the vector $\bm{w}$.
Evaluating the Hessian-vector products
$(\nabla^2_{\bm{x}\bm{x}} f) \bm{w}$, $(\nabla^2_{\bm{x}\bm{p}}f)\bm{w}$ and $(\nabla^2_{\bm{p}\bm{p}} f) \bm{w}$
is generally easier, as $f$ is a real-valued function.

\subsection{Second-order derivatives.}
\label{sec:reduction:ad}
To avoid forming the third-order tensors $\nabla^2 g$  in the reduction procedure
presented previously in \refsec{sec:reduction:soc}, we exploit
the particular structure of Equations~\refeqn{eq:socadjoint}
and \refeqn{eq:hessvecprod} to implement with automatic differentiation
an adjoint-tangent accumulation of the derivative information.
For any adjoint $\bm{\lambda} \in \mathbb{R}^{n_x}$ and vector $\bm{w} \in \mathbb{R}^{n_p}$, we build a tangent
$\bm{v} = (\bm{z}, \bm{w}) \in \mathbb{R}^{n_x + n_p}$, with $\bm{z} \in \mathbb{R}^{n_x}$
solution of the first system in Equation~\refeqn{eq:socadjoint}.
Then, the adjoint-forward accumulation evaluates a vector $\bm{y} \in \mathbb{R}^{n_x + n_p}$
as
\begin{equation}
  \label{eq:ADreduction}
  \bm{y} =  \begin{pmatrix}
   \bm{\lambda}^\top\nabla^2_{\bm{x}\bm{x}} g & \bm{\lambda}^\top\nabla^2_{\bm{x}\bm{p}} g \\
   \bm{\lambda}^\top\nabla^2_{\bm{p}{x}} g & \bm{\lambda}^\top\nabla^2_{\bm{p}\bm{p}} g
  \end{pmatrix}
  \bm{v} \; ,
\end{equation}
(the tensor projection notation will be introduced more thoroughly
in \refsec{sec:so:ad}).
We detail next how to compute the vector $\bm{y}$ by
using forward-over-reverse AutoDiff.

\subsubsection{AutoDiff.}
\label{sec:autodiff}
AutoDiff transforms a code that implements a multivariate vector function $\bm{y} = g(\bm{x}),\, \REAL^n \mapsto \REAL^m$ with inputs $\bm{x}$ and outputs $\bm{y}$ into its differentiated implementation. We distinguish  two modes of AutoDiff. Applying AutoDiff in {\it forward mode} generates the code for evaluating the Jacobian vector product $\bm{y}^{(1)} = \nabla g(\bm{x}) \cdot \bm{x}^{(1)}$, with the superscript $^{(1)}$ denoting first-order
tangents---also known as directional derivatives.  The {\it adjoint or reverse mode}, or backpropagation in machine learning, generates the code of the transposed Jacobian vector product $\bm{x}_{(1)} = \bm{y}_{(1)}\cdot \nabla g(\bm{x})^T$, with the subscript $_{(1)}$ denoting first-order adjoints. The adjoint mode is useful for computing gradients of scalar functions ($m=1$) (such as Lagrangian) at a cost of $\bigo{cost(g)}$.

\subsubsection{Sparse Jacobian Accumulation.}
To extract the full Jacobian from a tangent or adjoint AutoDiff
implementation, we have to let $\bm{x}^{(1)}$ and $\bm{y}_{(1)}$ go over the
Cartesian basis of $\REAL^n$ and $\REAL^m$, respectively. This incurs the
difference in cost for the Jacobian accumulation: $\bigo{n} \cdot cost(g)$
for the tangent Jacobian model and $\bigo{m} \cdot cost(g)$ for the adjoint
Jacobian model. In our case we need the full square ($m=n$) Jacobian $\nabla_{\bm{x}} g$
of the nonlinear function~\refeqn{eq:nonlinearsystem} to run the Newton--Raphson
algorithm. The tangent model is preferred whenever $m \approx n$.
Indeed, the adjoint model incurs a complete
reversal of the control flow and thus requires storing intermediate variables,
leading to high cost in memory.
Furthermore, SIMD architectures are particularly well suited for propagating
the $n$ independent tangent Jacobian vector products in parallel \cite{revels2018-ixedmode}.

\begin{figure}
  \centering
\includegraphics[width=0.7\linewidth]{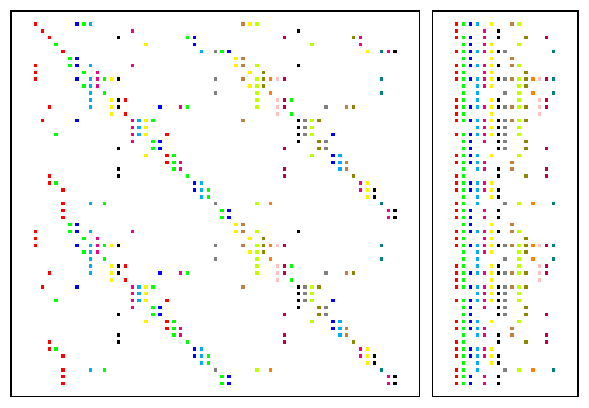}
\caption{Jacobian compression via column coloring. On the left, the original Jacobian.
On the right, the compressed Jacobian.}
\label{fig:coloring}
\end{figure}

If $n$ becomes larger (>>1000), however, the memory requirement of all $n$
tangents may exceed the GPU's memory. Since our Jacobian is sparse, we apply
the technique of Jacobian coloring that compresses independent columns of
the Jacobian and reduces the number of required {\it seeding} tangent vectors
from $n$ to the number of colors~$c$ (see Figure~\ref{fig:coloring}).

\subsubsection{Second-Order Derivatives.}
\label{sec:so:ad}
\begin{figure}
  \centering
\includegraphics[width=.8\linewidth]{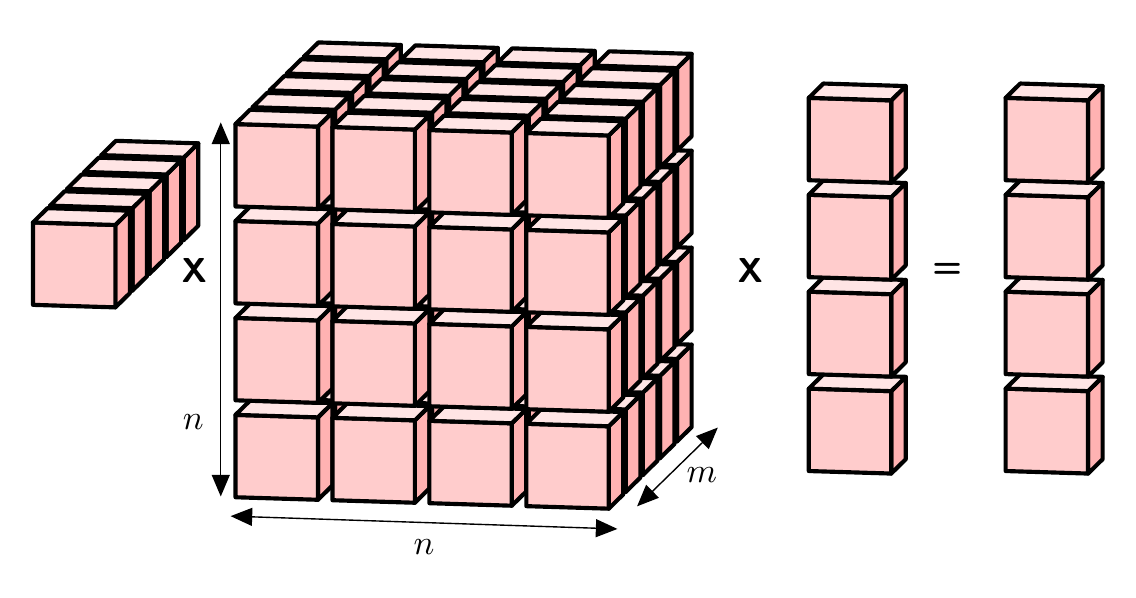}
\caption{Hessian derivative tensor projection $<\bm{y}_{(1)}, \nabla^2 g(\bm{x}), \bm{x}^{(2)}>$. Notice that the Hessian slices along the $n$ directions are symmetric.}
\label{fig:hessianprojection}
\end{figure}

For higher-order derivatives that involve derivative tensors (e.g.,
Hessian $\nabla^2 g \in \REAL^{m \times n \times n}$) we introduce the projection notation $<\cdots>$ introduced in
\cite{naumann2012art} and illustrated in \reffig{fig:hessianprojection} with
$<\bm{x}_{(1)}, \nabla^2 g(\bm{x}), \bm{x}^{(1)}>$, whereby adjoints are
projected from the left to the Jacobian and tangents from the right.
To compute second-order derivatives and the Hessian projections in
Equation~\refeqn{eq:ADreduction}, we use the adjoint model implementation given by
\begin{equation}
  \bm{y}=g(\bm{x}),\
  \bm{x}_{(1)} = < \bm{y}^{(1)}, \nabla g(\bm{x})> = \bm{y}_{(1)} \cdot \nabla
  g(\bm{x})^T  \,,
  \label{eq:adjoint}
\end{equation}
and we apply over it the tangent model given by
\begin{equation}
\begin{array}{llll}
  \bm{y}=g(\bm{x}), &
  \bm{y}^{(1)} = < \nabla g(\bm{x}), \bm{x}^{(1)}> = \nabla
  g(\bm{x}) \cdot \bm{x}^{(1)} \,,
  \label{eq:tlm}
\end{array}
\end{equation}
yielding
\begin{equation}
\begin{array}{llll}
  \bm{y}&=g(\bm{x}), \\
  \bm{y}^{(2)}&=< \nabla g(\bm{x}), \bm{x}^{(2)}>, \\
  \text{and} \\
  \bm{x}_{(1)}& = < \bm{y}_{(1)}, \nabla g(\bm{x})>, \\
  \bm{x}^{(2)}_{(1)} &= <\bm{y}_{(1)}, \nabla^2 g(\bm{x}), \bm{x}^{(2)}> +
  <\bm{y}^{(2)}_{(1)}, \nabla g(\bm{x})>\,.
\end{array}
\label{eq:so_model}
\end{equation}
Notice that every variable has now a value component and three
derivative components denoted by $_{(1)}$, $^{(2)}$, and $^{(2)}_{(1)}$
amounting to first-order adjoint, second-order tangent, and second-order
tangent over adjoint, respectively. In  \refsec{sec:reduction:algorithm}, we  compute
the term $x_{(1)}^{(2)}$ on the GPU
by setting $\bm{y}_{(1)}^{(2)} = 0$ and extracting the result from $\bm{x}_{(1)}^{(2)} \in \REAL^n$.

\subsection{Reduction Algorithm.}
\label{sec:reduction:algorithm}
We are now able to write down the reduction algorithm to compute the
Hessian-vector products $\nabla^2 F \cdot \bm{w}$. We first present a sequential
version of the algorithm, and then detail how to design a parallel variant
of the reduction algorithm.

\begin{algorithm2e}
  \KwData{Vector $\bm{w} \in \mathbb{R}^{n_p}$}
  {\tt SpMul}: $\bm{b} = \big(\nabla_{\bm{u}}g  \big) \bm{w}$ \;
  {\tt SparseSolve}: $(\grad{g}{\bm{x}}) \bm{z} = - \bm{b}$ \;
  {\tt TensorProjection}: Compute $(\bm{y}_{\bm{x}}, \bm{y}_{\bm{p}})$ with~\refeqn{eq:ADreduction} and $\bm{v} = (\bm{z}, \bm{w})$\;
  {\tt SparseSolve}: $(\grad{g}{\bm{x}})^\top \bm{\psi} = - \bm{y}_{\bm{x}}$ \;
  {\tt MulAdd}: $(\nabla^2 F)\bm{w} = \bm{y}_{\bm{p}} + (\grad{g}{\bm{p}})^\top\bm{\psi}$ \;
 \caption{Reduction algorithm}
 \label{algo:reduction}
\end{algorithm2e}

\subsubsection{Sequential algorithm.}
We observe that by default the Hessian reduction algorithm encompasses four sequential steps:
\begin{enumerate}
  \item {\tt SparseSolve}: Get the second-order adjoint $\bm{z}$
    by solving the first linear system in~\refeqn{eq:socadjoint}.
  \item {\tt TensorProjection}: Define the tangent $\bm{v} := (\bm{z}, \bm{w})$, and evaluate the
    second-order derivatives using~\refeqn{eq:ADreduction}. {\tt TensorProjection}
    returns a vector $\bm{y} = (\bm{y}_{\bm{x}}, \bm{y}_{\bm{p}})$,
    with
    \begin{equation}
      \label{eq:reduction}
      \left\{
        \begin{aligned}
          \bm{y}_{\bm{x}} =&
         <\bm{\lambda}^\top,\nabla_{\bm{x}\bm{x}}^2g,\bm{z}>+
         <\bm{\lambda}^\top,\nabla_{\bm{x}\bm{p}}^2g,\bm{w}>+ \\
         &<\nabla_{\bm{x}\bm{x}}^2f,\bm{z}>+
         <\nabla_{\bm{x}\bm{p}}^2f,\bm{w}>\; ,\\
          \bm{y}_{\bm{p}} =&
         <\bm{\lambda}^\top,\nabla_{\bm{p}\bm{x}}^2g,\bm{z}>+
         <\bm{\lambda}^\top,\nabla_{\bm{p}\bm{p}}^2g,\bm{w}>+\\
         &<\nabla_{\bm{p}\bm{x}}^2f,\bm{z}>+
         <\nabla_{\bm{p}\bm{p}}^2f,\bm{w}>\; ,
        \end{aligned}
      \right.
    \end{equation}
    with ``$<>$" denoting the derivative tensor projection introduced in \refsec{sec:so:ad}
    (and illustrated in \reffig{fig:hessianprojection}).
  \item {\tt SparseSolve}: Get the second-order adjoint $\bm{\psi}$ by solving the second
    linear system in Equation~\refeqn{eq:socadjoint}: $
    (\grad{g}{\bm{x}})^\top \bm{\psi} = - \bm{y}_{\bm{x}}$.
  \item {\tt SpMulAdd}: Compute the reduced Hessian-vector product with Equation~\refeqn{eq:hessvecprod}.
\end{enumerate}
The first {\tt SparseSolve} differs from the second {\tt SparseSolve} since the left-hand side is different: the first
system considers the Jacobian matrix $(\nabla_{\bm{x}}g)$, whereas the second system considers
its transpose $(\nabla_{\bm{x}}g)^\top$.

To compute the entire reduced
Hessian $\nabla^2 F$, we have to let $\bm{w}$ go over all the Cartesian basis vectors of $\REAL^{n_p}$.
The parallelization over these basis vectors is explained in the next paragraph.

\subsubsection{Parallel Algorithm.}
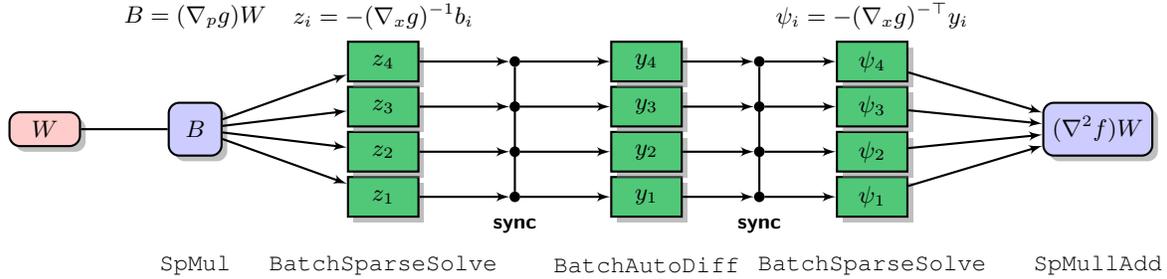
\begin{figure*}
    \centering
    \begin{tikzpicture}[font=\small\sffamily\bfseries,thick,auto, yscale=0.3]
      \node [input] (state) at (-2, 5) {$W$};
      \node [seq] (spmul) at (0, 5) {$B$};
      \node [seq] (spmul2) at (12, 5) {$(\nabla^2 f)W$};

        \path [draw] (state) -- (spmul);
        \foreach \x in {1,...,4}
        {
          \node [parallel]  (\x1) at (2.5,2*\x) {$z_{\x}$};
          \path [line] (spmul) -- (\x1);

          \node [n] (\x2) at (4.25, 2*\x) {};
          \path [line] (\x1) -- (\x2);

          \node [parallel]  (\x3) at (6,2*\x) {$y_{\x}$};
          \path [line] (\x2) -- (\x3);

          \node [n] (\x4) at (7.5, 2*\x) {};
          \path [line] (\x3) -- (\x4);

          \node [parallel] (\x5) at (9, 2*\x) {$\psi_{\x}$};
          \path [line] (\x4) -- (\x5);

          \path [line] (\x5) -- (spmul2);
        }

        \path [draw] (12) -- (42);
        \path [draw] (14) -- (44);

        \node (sync1) at (4.25, 0.7) {{\footnotesize sync}};
        \node (sync2) at (7.5, 0.7) {{\footnotesize sync}};
        \node (l11) at (0, -1) {{\tt SpMul}};
        \node (l12) at (0, 10) {$B = (\nabla_p g)W$};

        \node (l21) at (2.5, -1) {{\tt BatchSparseSolve}};
        \node (l22) at (2.5, 10) {$z_i = -(\nabla_x g)^{-1} b_i$};

        \node (l31) at (6, -1) {{\tt BatchAutoDiff}};

        \node (l41) at (9, -1) {{\tt BatchSparseSolve}};
        \node (l42) at (9, 10) {$\psi_i = -(\nabla_x g)^{-\top} y_i$};

        \node (l51) at (12, -1) {{\tt SpMullAdd}};
    \end{tikzpicture}

    \caption{Parallel computation of the reduced Hessian vector products on the GPU}
    \label{fig:reduction}
\end{figure*}

Instead of computing the Hessian vector products $(\nabla^2 F)\bm{w}_1,\cdots, (\nabla^2 F)\bm{w}_n$
one by one, the parallel algorithm takes as input a \emph{batch} of $N$ vectors $W = \big(\bm{w}_1, \cdots,
\bm{w}_N\big)$ and evaluates the Hessian-vector products
$\big((\nabla^2 F)\bm{w}_1, \cdots, (\nabla^2 F)\bm{w}_N\big)$ in a parallel fashion.
By replacing respectively the {\tt SparseSolve} and {\tt TensorProjection} blocks by
{\tt BatchSparseSolve} and {\tt BatchTensorProjection},  we get the
parallel reduction algorithm presented in Algorithm~\ref{algo:batchreduction}
(and illustrated in Figure~\ref{fig:reduction}).
On the contrary to Algorithm~\ref{algo:reduction}, the block
{\tt BatchSparseSolve} solves a sparse linear system with multiple right-hand-sides
$B = (\nabla_{\bm{p}}g) W$, and the block {\tt BatchTensorProjection} runs the Autodiff
algorithm introduced in \refsec{sec:reduction:ad} in batch.
As explained in the next section,
both operations are fully amenable to the GPU.
\begin{algorithm2e}
  \KwData{$N$ vectors $\bm{w}_1, \cdots, \bm{w}_N \in \mathbb{R}^{n_p}$}
  Build $W = (\bm{w}_1, \cdots, \bm{w}_N)$ $, W \in \mathbb{R}^{n_p \times N}$  \;
  {\tt SpMul}: $B = \big(\nabla_{\bm{p}}g  \big) W$ $, B \in \mathbb{R}^{n_x \times N}$, $\nabla_{\bm{p}}g \in \mathbb{R}^{n_x \times n_p}$\;
  {\tt BatchSparseSolve}: $(\grad{g}{\bm{x}}) Z = - B$ \;
  {\tt BatchTensorProjection}: Compute $(Y_{\bm{x}}, Y_{\bm{p}})$ with $V = (Z, W)$ \;
  {\tt BatchSparseSolve}: $(\grad{g}{\bm{x}})^\top \Psi = - Y_{\bm{x}}$ \;
  {\tt SpMulAdd}: $(\nabla^2 F)\bm{W} = Y_{\bm{p}} + (\grad{g}{\bm{p}})^\top\Psi$ \;
 \caption{Parallel reduction algorithm}
 \label{algo:batchreduction}
\end{algorithm2e}

\section{GPU Implementation}
\label{sec:implementation}

In the previous section, we have devised a parallel algorithm to compute
the reduced Hessian. This algorithm involves two key ingredients, both
running in parallel: {\tt BatchSparseSolve} and {\tt BatchTensorProjection}.
We present in \refsec{sec:implementation:autodiff} how to implement
{\tt BatchTensorProjection} on GPU by leveraging the Julia language. Then,
we focus on the parallel
resolution of {\tt BatchSparseSolve} in \refsec{sec:implementation:linearalgebra}.
The final implementation is presented in \refsec{sec:implementation:parallel}.

\subsection{Batched AutoDiff.}
\label{sec:implementation:autodiff}

\subsubsection{AutoDiff on GPU.}
\label{sec:implementation:design}

Our implementation attempts to be architecture agnostic, and to this
end we rely heavily on the just-in-time compilation capabilities of the Julia language.
Julia has two key advantages for us: (i) it implements state-of-the-art automatic differentiation
libraries and (ii) its multiple dispatch capability allows to write code in
an architecture agnostic way.
Combined together, this allows to run AutoDiff on GPU accelerators.
On the architecture side we rely on the array
abstraction implemented by the package \lstinline{GPUArrays.jl}~\cite{besard2018effective} and on the
kernel abstraction layer \KA.
The Julia community provides three GPU backends for these two packages:
NVIDIA, AMD, and Intel oneAPI. Currently, \lstinline{CUDA.jl} is the most
mature package, and we are leveraging this infrastructure to run our code on an
x64/PPC CPU and NVIDIA GPU. In the future our solution will be rolled out
transparently onto AMD and Intel accelerators with minor code changes.

\subsubsection{Forward Evaluation of Sparse Jacobians.}
\label{sec:jacobian:ad}
The reduction algorithm in \refsec{sec:reduction:algorithm} requires (i) the Jacobian
$\nabla_x g$ to form the linear system in \refeqn{eq:socadjoint} and (ii)
the Hessian vector product of $\bm{\lambda}^\top \nabla^2 g$ in \refeqn{eq:reduction}.
We use the Julia package \lstinline{ForwardDiff.jl}~\cite{revels2016forward} to apply the first-order
tangent model \refeqn{eq:tlm} by instantiating every variable as a dual type
defined as \verb+T1S{T,C} = ForwardDiff.Dual{T, C}}+,
where \lstinline{T} is the type ({\tt double} or {\tt float}) and
\lstinline{C} is the number of directions that are propagated together in
parallel. This allows us to apply AutoDiff both on the CPU and on the GPU in a
vectorized fashion, through a simple type change:
for instance,
\lstinline+Array{TIS{T, C}}(undef, n)+ instantiates a vector of dual numbers on the CPU,
whereas \lstinline+CuArray{TIS{T, C}}(undef, n)+ does the same on a CUDA GPU.
(Julia allows us to write code where all the types are abstracted away).
This, combined with \KA, allows us to write a portable residual kernel for
$g(\bm{x}, \bm{p})$ that is both differentiable and architecture agnostic. By setting the
number of Jacobian colors $c$ to the parameter \lstinline|C| of type \lstinline|T1S{T,C}| we
leverage the GPUs by propagating the tangents in a SIMD way.

\subsubsection{Forward-over-Reverse Hessian Projections.}
\label{sec:hessian:ad}
As opposed to the forward mode,
generating efficient adjoint code for GPUs is known to be hard.
Indeed, adjoint automatic differentiation implies a reversal of the computational flow,
and in the backward pass every read of a variable translates to a write adjoint,
and vice versa. The latter is particularly complex for parallelized
algorithms, especially as the automatic parallelization of algorithms
is hard. For example, an
embarrassingly parallel algorithm where each process reads the data of all
the input space leads to a challenging race condition in its adjoint.
Current state-of-the-art AutoDiff tools use specialized
workarounds for certain cases. However, a generalized solution to this
problem does not exist. The promising AutoDiff tool Enzyme
\cite{enzymeNeurips} is able to differentiate CUDA kernels in Julia, but  it
is currently not able to digest all of our code.

To that end, we hand differentiate our GPU kernels for the
forward-over-reverse Hessian projection. We then apply
\lstinline{ForwardDiff} to these adjoint kernels to extract second-order
sensitivities according to the forward-over-reverse model.
Notably, our test case (see \refsec{sec:background:powerflow}) involves reversing a
graph-based problem (with vertices $V$ and edges $E$). The variables of the
equations are defined on the vertices. To adjoin or reverse these kernels, we
pre-accumulate the adjoints first on the edges and then on the nodes, thus
avoiding a race condition on the nodes. This process yields a fully
parallelizable adjoint kernel. Unfortunately, current AutoDiff tools are not
capable of detecting such structural properties. Outside the kernels we use a
tape (or stack) structure to store the values computed in the forward pass
and to reuse them in the reverse (split reversal). The kernels themselves are
written in joint reversal, meaning that the forward and reverse passes are
implemented in one function evaluation without intermediate storage of
variables in a data structure. For a more detailed introduction to writing
adjoint code we recommend \cite{griewank2008evaluating}.

\subsection{Batched Sparse Linear Algebra.}
\label{sec:implementation:linearalgebra}

The block {\tt BatchSparseSolve} presented in \refsec{sec:reduction:algorithm}
requires the resolution of two sparse linear systems with multiple right-hand sides,
as illustrated in Equation~\eqref{eq:socadjoint}.
This part is critical because in practice a majority of the time
is spent inside the linear algebra library in the parallel reduction algorithm.
To this end, we have wrapped the library
\cusolverrf\ in Julia to get an efficient LU solver on the GPU.
For any sparse matrix $A \in \mathbb{R}^{n \times n}$,
the library \cusolverrf\ takes as input an LU factorization of the matrix $A$ precomputed on the host,
and transfers it to the device. \cusolverrf\ has two key advantages to implement
the resolution of the two linear systems in {\tt BatchSparseSolve}.
(i) If a new matrix $\tilde A$ needs to be factorized and has the same
sparsity pattern as the original matrix $A$, the refactorization routine
proceeds directly on the device, without any data transfer with the host
(allowing to match the performance of the state-of-the-art CPU sparse library UMFPACK~\cite{davis2004algorithm}).
(ii) Once the LU factorization has been computed, the forward and backward solves
for different right-hand  sides $\bm{b}_1, \cdots, \bm{b}_N$ can be computed in batch mode.

\subsection{Implementation of the Parallel Reduction.}
\label{sec:implementation:parallel}

By combining  the batch AutoDiff with the batch sparse linear
solves of \cusolverrf, we  get a fully parallel algorithm to
compute the reduced Hessian projection. We compute the reduced
Hessian $\nabla^2 F \in \mathbb{R}^{n_p \times n_p}$ by blocks of $N$ Hessian-vector products.
If we have enough memory to set $N = n_p$, we can compute the full reduced
Hessian in one batch reduction. Otherwise, we set $N < n_p$ and compute
the full reduced Hessian in $N_b = div(n, N) + 1$ batch reductions.

Tuning the number of batch reductions $N$ is nontrivial and depends on
two considerations.
How efficient is the parallel scaling when we run the two parallel blocks {\tt BatchTensorProjection} and {\tt BatchSparseSolve}? and
 Are we fitting into the device memory?
This second consideration is indeed one of the bottlenecks of the algorithm.
In fact, if we look more closely at the memory usage of the parallel
reduced Hessian, we observe that the memory grows linearly with the number
of batches $N$. First, in the block {\tt BatchTensorProjection}, we need to duplicate $N$ times the
tape used in the reverse accumulation of the Hessian in \refsec{sec:implementation:autodiff},
leading to memory increase from $\mathcal{O}(M_T)$ to $\mathcal{O}(M_T \times N)$, with $M_T$ the memory of the tape.
The principle is similar in {\tt SparseSolve}, since the second-order
adjoints $\bm{z}$ and $\bm{\psi}$ are also duplicated in batch mode,
leading to a memory increase from $\mathcal{O}(2 n_x)$ to $\mathcal{O}(2 n_x  \times N)$.
This is a bottleneck on large cases when the number of variables $n_x$
is large.

The other bottleneck arises when we combine together the blocks {\tt BatchSparseSolve}
and {\tt BatchTensorProjection}. Indeed, {\tt BatchTensorProjection} should wait for the first
block {\tt BatchSparseSolve} to finish its operations. The same
issue arises when passing the results of {\tt BatchTensorProjection} to the second
{\tt BatchSparseSolve} block. As illustrated by Figure~\ref{fig:reduction},
we need to add two explicit synchronizations in the algorithm.
Allowing the algorithm to run the reduction algorithm in a purely asynchronous
fashion would require a tighter integration with \cusolverrf.

\section{Numerical experiments}
\label{sec:results}
In this section we  provide extensive benchmarking results that investigate
whether the computation of the reduced Hessian $\nabla^2 f$ with \refalg{algo:batchreduction}
is well suited for SIMD on GPU architectures.
As a comparison, we use a CPU implementation based on the sparse LU solver UMFPACK,
with iterative refinement disabled\footnote{We set the parameter {\tt UMFPACK\_IRSTEP} to 0.} (it yields no numerical
improvement, however, considerably speeds up the computation).
We show that on the largest instances our GPU implementation is 30 times faster than
its sequential CPU equivalent and provide a path forward to further improve our implementation.
Then, we illustrate that the reduced Hessian computed is effective to track a suboptimal in a real-time
setting.

\subsection{Experimental Setup}
\subsubsection{Hardware.}

Our workstation {\it Moonshot} is provided by Argonne National Laboratory.
All the experiments run on a NVIDIA V100 GPU (with 32GB of memory)
and {\tt CUDA 11.3}. The system is equipped with a Xeon Gold 6140, used to run the experiments on the CPU (for comparison).
For the software, the workstation works with Ubuntu 18.04,
and we use Julia 1.6 for our implementation.
We rely on our package \KA\ and \lstinline{GPUArrays.jl} to generate parallel GPU code.

All the implementation is open-sourced, and an artifact is provided to reproduce
the numerical results\footnote{available on \url{https://github.com/exanauts/Argos.jl/tree/master/papers/pp2022}}.

\subsubsection{Benchmark library.}
The test data represents various case instances (see
\reftab{tab:test_instances}) in the power grid community obtained
from the open-source benchmark library PGLIB~\cite{babaeinejadsarookolaee2019power}.
The number in the case name
indicates the number of buses (graph nodes) $n_v$ and the number of lines (graph edges) $n_e$ in the power grid: $n_x$ is
the number of variables, while $n_p$ is the number of parameters (which
is also equal to the dimension of the reduced Hessian and the parameter space $\REAL^{n_p}$).

\begin{table}[!ht]
  \centering
  \resizebox{.35\textwidth}{!}{
\begin{tabular}{l|cc|cc}
{\bf Case} & {\bf $n_v$} & {\bf $n_e$}& {\bf $n_x$} & {\bf $n_p$} \\
\hline
IEEE118 & 118 & 186 & 181 & 107 \\
IEEE300 & 300 & 411 & 530 & 137 \\
PEGASE1354 & 1,354 & 1,991 & 2,447 & 519 \\
PEGASE2869 & 2,869 & 4,582 & 5,227 & 1,019 \\
PEGASE9241 & 9,241 & 16,049 & 17,036 & 2,889 \\
GO30000 & 30,000 & 35,393 & 57,721 & 4,555
\end{tabular}
}
\caption{Case instances obtained from PGLIB}
\label{tab:test_instances}
\end{table}

\subsection{Numerical Results}

\subsubsection{Benchmark reduced Hessian evaluation.}
\begin{table}[!ht]
  \centering
  \resizebox{.44\textwidth}{!}{
    \begin{tabular}{l|rrl}
      \multirow{2}{*}{\bf Cases}& \multicolumn{3}{c}{{\bf Dimensions}}\\
                                & $W \in \mathbb{R}^{n_p \times N}$ & $B\in \mathbb{R}^{n_x \times N}$ & $\nabla^2 f \in \mathbb{R}^{n_p \times n_p}$\\
                                \hline
      IEEE118 & $107 \times N $ & $181 \times N$ & $107 \times 107$\\
      IEEE300 & $137 \times N $ & $530 \times N$ & $137 \times 137$\\
      PEGASE1354 & $519 \times N $ & $2,447 \times N$ & $519 \times 519$\\
      PEGASE2869 & $1,019 \times N $ & $5,227 \times N$ & $1,019 \times 1,019$\\
      PEGASE9241 & $17,036 \times N $ & $17,036 \times N$ & $2,889 \times 2,889$\\
      GO30000 & $30,000 \times N$ & $35,393 \times N$ & $4,555 \times 4,555$
    \end{tabular}
  }
  \caption{Size of key matrices (seed matrix $W$, multiple right-hand sides
    $B$, and final reduced Hessian $\nabla^2 F$) for a batch size of $N$.
    On GO30000, instantiating the three
    matrices $W, B, \nabla^2 F$ for $N=256$ already takes 286MB in the GPU memory.
  }
  \label{tab:test_instances_size}.
\end{table}

For the various problems described in \reftab{tab:test_instances}, we
benchmarked the computation of the reduced Hessian $\nabla^2 F$ for different
batch sizes $N$. Each batch computes $N$ columns of the reduced Hessian (which
has a fixed size of $n_p \times n_p$).
Hence, the algorithm requires $N_b = div( n_p, N) +1$
number of batches to evaluate the full Hessian.

\begin{figure}[!ht]
    \centering
    \includegraphics[width=\linewidth]{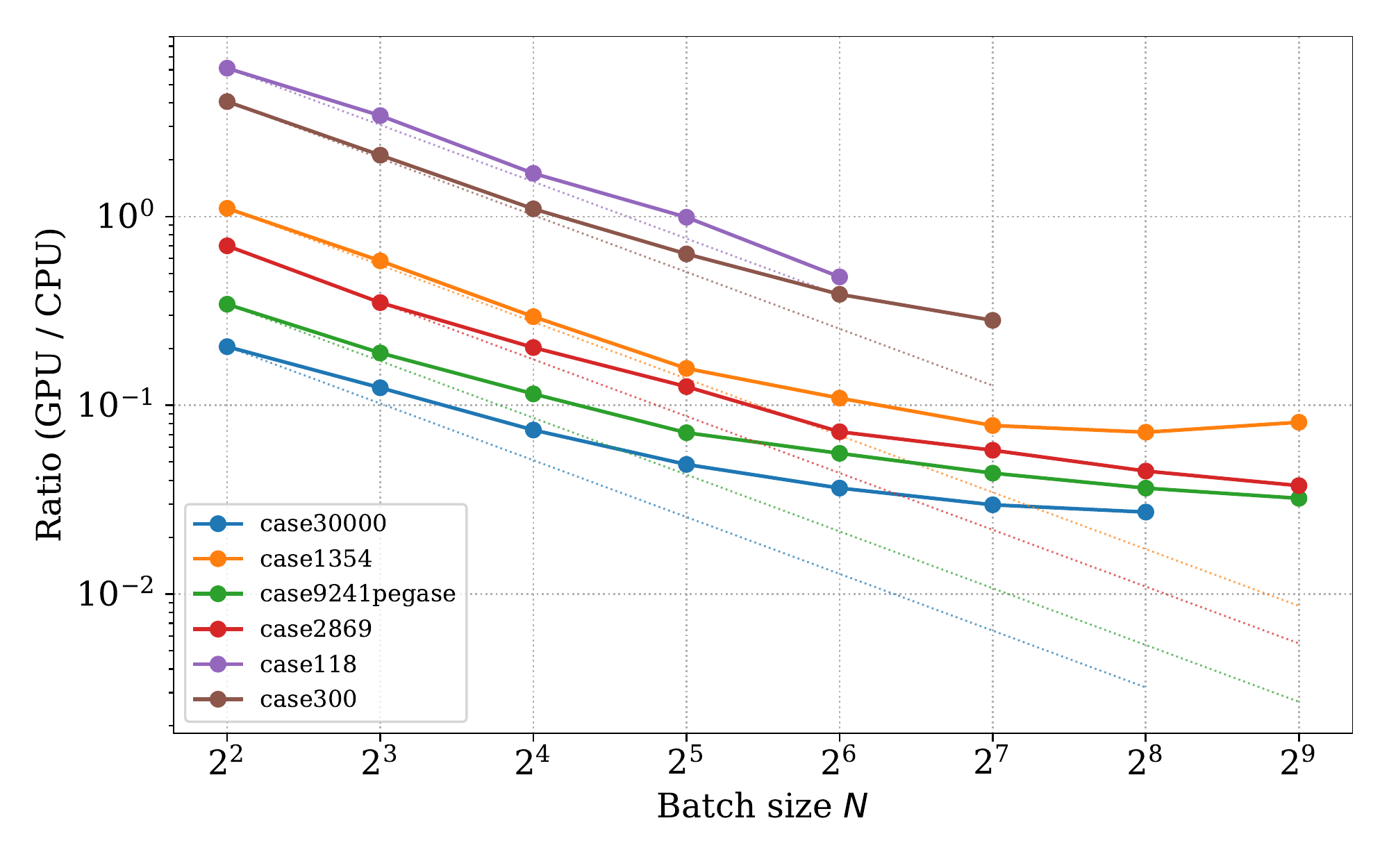}
    \caption{
      Parallel scaling of the total reduced Hessian accumulation $\nabla^2 F$
      with batch size $N$: A ratio value $< 1$ indicates a faster runtime
      compared with that of UMFPACK and AutoDiff on the CPU in absolute time. The
      dotted lines indicate the linear scaling reference. Lower values imply a
      higher computational intensity.
    }
    \label{fig:batch_scaling}
\end{figure}

In Figure~\ref{fig:batch_scaling}, we compare on various instances (see
\reftab{tab:test_instances_size}) the reference CPU implementation together
with the full reduced Hessian computation $\nabla^2 F$ on the GPU
(with various batch sizes $N$). The figure is displayed in log-log scale, to better illustrate the linear scaling of the algorithm.
In addition, we scale the time taken by
the algorithm on the GPU by the time taken
to compute the full reduced Hessian on the CPU: a value below 1 means that the GPU is faster than the CPU.

We observe that the larger the number of batches $N$, the faster the GPU implementation is. This proves that the GPU is effective at parallelizing the reduction algorithm, with a scaling almost linear
when the number of batches is small ($N < 32 = 2^5$).  However, we reach the scalability limit of the GPU as we increase the number of batches $N$ (generally, when $N \geq 256= 2^8$).
Comparing to the CPU implementation, the speed-up is not large on small instances ($\approx$ 2 for IEEE118 and
IEEE300), but we get up to a 30 times speed-up on the largest instance GO30000, when using a
large number of batches.

\begin{figure}[!ht]
    \centering
    \includegraphics[width=\linewidth]{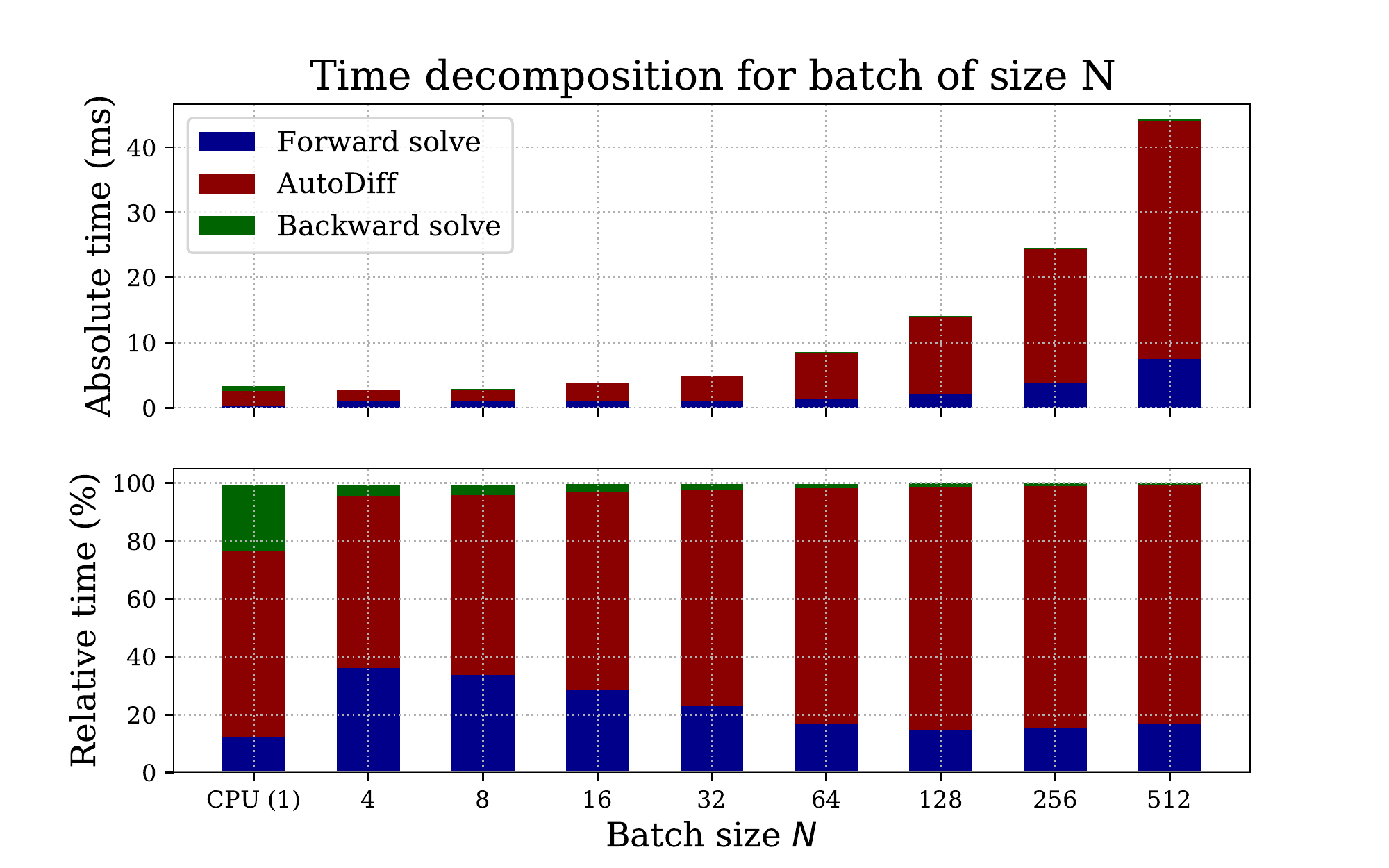}
    \caption{
      Decomposition of the runtime against the number of
      batch $N$, on case PEGASE 9241. $N=1$ corresponds to the CPU implementation.
      The derivative
      computation is the dominant kernel.}
    \label{fig:batch_time_decomposed}
\end{figure}

\reffig{fig:batch_time_decomposed} shows the relative time spent in the linear algebra and the automatic differentiation backend.
On the CPU, we observe that UMFPACK is very efficient to perform the linear solves (once the iterative
refinement is deactivated). However, a significant amount of the total running time is spent inside the AutoDiff
kernel. We get a similar behavior on the GPU: the batched automatic differentiation backend leads
to a smaller speed-up than the linear solves, increasing the fraction of the total runtime
spent in the block {\tt BatchAutoDiff}.

\subsubsection{Discussion.}

Our analysis shows that the reduced Hessian scales with the batch size, while
hitting an utilization limit for larger test cases. Our kernels may still have potential for improvement, thus further improving utilization scaling as long as we do not hit the memory capacity limit.
However, the sparsity of the power flow problems represents a worst-case problem for SIMD architectures, common
in graph-structured applications. Indeed, in contrast to PDE-structured
problems, graphs are difficult to handle in SIMD architectures because
of their unstructured sparsity patterns.


\subsection{Real-time tracking algorithm.}
Finally, we illustrate the benefits of our reduced Hessian algorithm by embedding it in a real-time
tracking algorithm.

Let $\bm{w}_t = (\bm{P}^d_t, \bm{Q}_t^d)$ be the loads
in~\eqref{eq:powerflowvec}, indexed by time $t$ and updated every minute.
In that setting, the reduced space problem is parameterized by the loads~$\bm{w}_t$:
\begin{equation}
  \label{eq:nonlinearoptreduced_time}
  \min_{\bm{p}_t} \; F(\bm p_t; \bm w_t) := f\big(x(\bm{p}_t), \bm{p}_t; \bm{w}_t\big) \; .
\end{equation}
For all time $t$, the real-time algorithm aims at tracking the optimal solutions $\bm{p}_t^\star$
associated with the sequence of problems \eqref{eq:nonlinearoptreduced_time}. To achieve this,
we update the tracking point $\bm{p}_t$
at every minute, by exploiting the curvature information provided by the reduced Hessian. The procedure is the following:
\begin{itemize}
  \item \textbf{Step 1:} For new loads $\bm{w}_t = (\bm{P}_t^d, \bm{Q}_t^d)$, compute
    the reduced gradient $\bm{g}_t = \nabla_{\bm{p}} F(\bm{p}_t; \bm{w}_t)$
    and the reduced Hessian $H_t = \nabla^2_{\bm{p}\bm{p}} F(\bm{p}_t; \bm{w}_t)$
    using Algorithm~\ref{algo:batchreduction}.
  \item \textbf{Step 2:} Update the tracking control $\bm{p}_t$ with
    $\bm{p}_{t+1} = \bm{p}_t + \bm{d}_t$, where $\bm{d}_t$
    is a descent direction computed as solution of the dense linear system
    \begin{equation}
      \label{eq:qp_rto}
      H_t \; \bm{d}_t = - \bm{g}_t \; .
    \end{equation}
\end{itemize}
In practice, we use the dense Cholesky factorization implemented in \cusolver\
to solve the dense linear system~\eqref{eq:qp_rto} efficiently on the GPU.

We compare the tracking controls $\{\bm{p}_t\}_{t=1,\cdots,T}$ with the optimal solutions
$\{\bm{p}_t^\star\}_{t=1,\cdots,T}$ associated to the sequence of optimization problems~\eqref{eq:nonlinearoptreduced_time}.
Note that solving each~\eqref{eq:nonlinearoptreduced_time} to optimality is an expensive operation, involving
calling a nonlinear optimization solver. On the contrary, the real-time tracking algorithm
involves only (i) updating the gradient and the Hessian for the new loads $\bm{w}_t$
and (ii) solving the dense linear system~\eqref{eq:qp_rto}.

\begin{figure}[!ht]
    \centering
    \includegraphics[width=\linewidth]{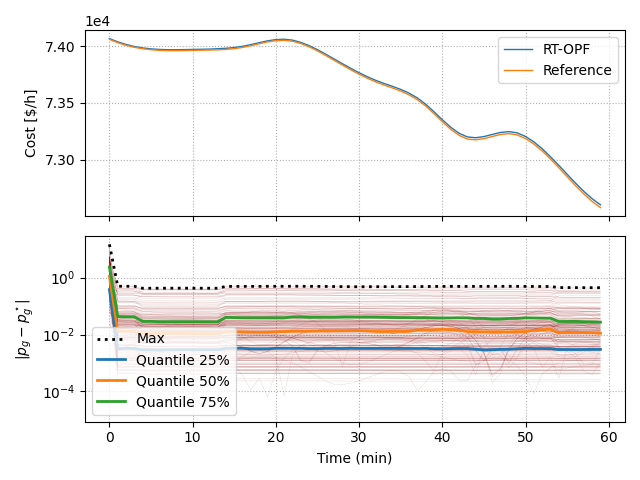}
    \caption{Performance of the real-time tracking algorithm on PEGASE1354, compared
      with the optimal solutions. The real-time algorithm is applied every minute, during one hour.
      The first plot shows the evolution
      of the operating cost along time, whereas the second plot shows the evolution
      of the absolute difference between the tracking control $\bm{p}_t$ and the optimum
    $\bm{p}_t^\star$.}
    \label{fig:rto}
\end{figure}
We depict in Figure~\ref{fig:rto} the performance of the real-time tracking algorithm, compared
with an optimal solution computed by a nonlinear optimization solver.
In the first subplot, we observe that the operating cost associated
to $\{\bm{p}_t\}_t$ is close to the optimal cost associated to $\{\bm{p}_t^\star\}_t$.
The second subplot depicts the evolution of the absolute difference $| \bm{p}_t - \bm{p}_t^\star|$,
component by component. We observe that the difference remains tractable: the median
(Quantile 50\%) is almost constant, and close to $10^{-2}$ (which in our case is not a large deviation
from the optimum) whereas the maximum difference remains below $0.5$. At each time $t$, the real-time
algorithm takes in average $0.10$s to update $\bm{p}_t$ on the GPU (with $N=256$ batches), comparing
to $2.22$s on the CPU (see Table~\ref{tab:timing_rto}). We achieve such a 20 times speed-up on the GPU
as
(i) the evaluation of the reduced Hessian is faster on the GPU
(ii) we do not have any data transfer between the host and the device to perform
the dense Cholesky factorization with {\tt cusolver}.
Hence, this real-time use case leverages the high parallelism of our algorithm
to evaluate the reduced Hessian.
\begin{table}[!ht]
  \centering
  \resizebox{.4\textwidth}{!}{
    \begin{tabular}{l|ccr}
       \hline
       & Step 1 (s) & Step 2 (s) & Total (s) \\
       \hline
      CPU & 1.41 & 0.81 & 2.22  \\
      GPU & 0.05 & 0.05 & 0.10 \\
       \hline
    \end{tabular}
  }
    \caption{Time to update the tracking point $\bm{p}_t$ for {\tt case1354pegase} with the real-time algorithm,
      on the CPU and on the GPU.
    }
    \label{tab:timing_rto}
\end{table}

\section{Conclusion}
In this paper we have devised and implemented a practical batched algorithm (see \refalg{algo:batchreduction})
to extract on SIMD architectures the second-order sensitivities from a system
of nonlinear equations. Our implementation on NVIDIA
GPUs leverages the programming language Julia to generate portable kernels
and differentiated code. We have observed that on the largest cases the batch
code is 30x faster than a reference CPU implementation using
UMFPACK. This is important for upcoming large-scale computer systems where availability of general purpose CPUs is very limited.
We have illustrated the interest of the reduced Hessian when used inside
a real-time tracking algorithm.
Our solution adheres to the paradigm of differential and composable
programming, leveraging the built-in metaprogramming capabilities of Julia.
In the future, we will investigate extending the method
to other classes of problems (such as uncertainty
quantification, optimal control, trajectory optimization, or PDE-constrained optimization).

\section*{Acknowledgments}
This research was supported by the Exascale Computing Project (17-SC-20-SC),
a joint project of the U.S. Department of Energy’s Office of Science and
National Nuclear Security Administration, responsible for delivering a
capable exascale ecosystem, including software, applications, and hardware
technology, to support the nation’s exascale computing imperative.

\small
\bibliographystyle{abbrv}
\bibliography{abbrv,paper.bib}

\begin{thebibliography}{10}

\bibitem{tensorflow2015-whitepaper}
M.~Abadi, A.~Agarwal, P.~Barham, E.~Brevdo, Z.~Chen, C.~Citro, G.~S. Corrado,
  A.~Davis, J.~Dean, M.~Devin, S.~Ghemawat, I.~Goodfellow, A.~Harp, G.~Irving,
  M.~Isard, Y.~Jia, R.~Jozefowicz, L.~Kaiser, M.~Kudlur, J.~Levenberg,
  D.~Man\'{e}, R.~Monga, S.~Moore, D.~Murray, C.~Olah, M.~Schuster, J.~Shlens,
  B.~Steiner, I.~Sutskever, K.~Talwar, P.~Tucker, V.~Vanhoucke, V.~Vasudevan,
  F.~Vi\'{e}gas, O.~Vinyals, P.~Warden, M.~Wattenberg, M.~Wicke, Y.~Yu, and
  X.~Zheng.
\newblock {TensorFlow}: Large-scale machine learning on heterogeneous systems,
  2015.

\bibitem{ahmadi2018parallel}
A.~Ahmadi, S.~Jin, M.~C. Smith, E.~R. Collins, and A.~Goudarzi.
\newblock Parallel power flow based on openmp.
\newblock In {\em 2018 North American Power Symposium (NAPS)}, pages 1--6.
  IEEE, 2018.

\bibitem{babaeinejadsarookolaee2019power}
S.~Babaeinejadsarookolaee, A.~Birchfield, R.~D. Christie, C.~Coffrin,
  C.~DeMarco, R.~Diao, M.~Ferris, S.~Fliscounakis, S.~Greene, R.~Huang, et~al.
\newblock The power grid library for benchmarking {AC} optimal power flow
  algorithms.
\newblock {\em arXiv preprint arXiv:1908.02788}, 2019.

\bibitem{beda1959programs}
L.~Beda et~al.
\newblock Programs for automatic differentiation for the machine besm.
\newblock {\em Precise Mechanics and Computation Techniques, Academy of
  Science, Moscow}, 1959.

\bibitem{besard2018effective}
T.~Besard, C.~Foket, and B.~De~Sutter.
\newblock Effective extensible programming: unleashing {Julia on GPUs}.
\newblock {\em IEEE Transactions on Parallel and Distributed Systems},
  30(4):827--841, 2018.

\bibitem{bezanson2017julia}
J.~Bezanson, A.~Edelman, S.~Karpinski, and V.~B. Shah.
\newblock Julia: A fresh approach to numerical computing.
\newblock {\em SIAM Rev.}, 59(1):65--98, 2017.

\bibitem{biegler2003large}
L.~T. Biegler, O.~Ghattas, M.~Heinkenschloss, and B.~van Bloemen~Waanders.
\newblock Large-scale {PDE}-constrained optimization: an introduction.
\newblock In {\em Large-Scale PDE-Constrained Optimization}, pages 3--13.
  Springer, 2003.

\bibitem{bluhdorn2020automat}
J.~Bl{\"u}hdorn, N.~R. Gauger, and M.~Kabel.
\newblock {AutoMat} -- automatic differentiation for generalized standard
  materials on gpus.
\newblock {\em arXiv preprint arXiv:2006.04391}, 2020.

\bibitem{bryson1962steepest}
A.~E. Bryson and W.~F. Denham.
\newblock A steepest-ascent method for solving optimum programming problems.
\newblock {\em Journal of Applied Mechanics}, 29:247, 1962.

\bibitem{davis2004algorithm}
T.~A. Davis.
\newblock Algorithm 832: {UMFPACK} v4.3---an unsymmetric-pattern multifrontal
  method.
\newblock {\em ACM Trans. Math. Software}, 30(2):196--199, 2004.

\bibitem{dommel1968optimal}
H.~W. Dommel and W.~F. Tinney.
\newblock Optimal power flow solutions.
\newblock {\em IEEE Transactions on power apparatus and systems},
  (10):1866--1876, 1968.

\bibitem{gilbert1992automatic}
J.~C. Gilbert.
\newblock Automatic differentiation and iterative processes.
\newblock {\em Optimization methods and software}, 1(1):13--21, 1992.

\bibitem{grabner2008automatic}
M.~Grabner, T.~Pock, T.~Gross, and B.~Kainz.
\newblock Automatic differentiation for {GPU-accelerated} 2d/3d registration.
\newblock In {\em Advances in automatic differentiation}, pages 259--269.
  Springer, 2008.

\bibitem{griewank2008evaluating}
A.~Griewank and A.~Walther.
\newblock {\em Evaluating derivatives: principles and techniques of algorithmic
  differentiation}.
\newblock SIAM, 2008.

\bibitem{huckelheim2018parallelizable}
J.~C. H{\"u}ckelheim, P.~D. Hovland, M.~M. Strout, and J.-D. M{\"u}ller.
\newblock Parallelizable adjoint stencil computations using transposed
  forward-mode algorithmic differentiation.
\newblock {\em Optimization Methods and Software}, 33(4-6):672--693, 2018.

\bibitem{innes2018flux}
M.~Innes.
\newblock Flux: Elegant machine learning with julia.
\newblock {\em Journal of Open Source Software}, 3(25):602, 2018.

\bibitem{kardos2020reduced}
J.~Kardos, D.~Kourounis, and O.~Schenk.
\newblock Reduced-space interior point methods in power grid problems.
\newblock {\em arXiv preprint arXiv:2001.10815}, 2020.

\bibitem{linnainmaa1976taylor}
S.~Linnainmaa.
\newblock Taylor expansion of the accumulated rounding error.
\newblock {\em BIT Numer. Math.}, 16(2):146--160, 1976.

\bibitem{enzymeNeurips}
W.~S. Moses and V.~Churavy.
\newblock Instead of rewriting foreign code for machine learning, automatically
  synthesize fast gradients.
\newblock In {\em Advances in Neural Information Processing Systems},
  volume~33, pages 12472--12485. 2020.

\bibitem{10.1145/3458817.3476165}
W.~S. Moses, V.~Churavy, L.~Paehler, J.~H\"{u}ckelheim, S.~H.~K. Narayanan,
  M.~Schanen, and J.~Doerfert.
\newblock Reverse-mode automatic differentiation and optimization of {GPU}
  kernels via {Enzyme}.
\newblock In {\em Proceedings of the International Conference for High
  Performance Computing, Networking, Storage and Analysis}, SC '21, New York,
  NY, USA, 2021. Association for Computing Machinery.

\bibitem{naumann2012art}
U.~Naumann.
\newblock {\em The Art of Differentiating Computer Programs: An Introduction to
  Algorithmic Differentiation}.
\newblock SIAM, 2012.

\bibitem{nolan1953analytical}
J.~F. Nolan.
\newblock {\em Analytical differentiation on a digital computer}.
\newblock PhD thesis, Massachusetts Institute of Technology, 1953.

\bibitem{papadimitriou2008direct}
D.~Papadimitriou and K.~Giannakoglou.
\newblock Direct, adjoint and mixed approaches for the computation of hessian
  in airfoil design problems.
\newblock {\em International journal for numerical methods in fluids},
  56(10):1929--1943, 2008.

\bibitem{NEURIPS2019_9015}
A.~Paszke et~al.
\newblock {PyTorch}: An imperative style, high-performance deep learning
  library.
\newblock In {\em Advances in Neural Information Processing Systems 32}, pages
  8024--8035. Curran Associates, Inc., 2019.

\bibitem{pearlmutter1994fast}
B.~A. Pearlmutter.
\newblock Fast exact multiplication by the hessian.
\newblock {\em Neural computation}, 6(1):147--160, 1994.

\bibitem{revels2018-ixedmode}
J.~Revels, T.~Besard, V.~Churavy, B.~D. Sutter, and J.~P. Vielma.
\newblock Dynamic automatic differentiation of {GPU} broadcast kernels.
\newblock {\em CoRR}, abs/1810.08297, 2018.

\bibitem{revels2016forward}
J.~Revels, M.~Lubin, and T.~Papamarkou.
\newblock Forward-mode automatic differentiation in {Julia}.
\newblock {\em arXiv preprint arXiv:1607.07892}, 2016.

\bibitem{tang2017real}
Y.~Tang, K.~Dvijotham, and S.~Low.
\newblock Real-time optimal power flow.
\newblock {\em IEEE Transactions on Smart Grid}, 8(6):2963--2973, 2017.

\bibitem{tinney1967power}
W.~F. Tinney and C.~E. Hart.
\newblock Power flow solution by {Newton's} method.
\newblock {\em IEEE Transactions on Power Apparatus and systems},
  (11):1449--1460, 1967.

\bibitem{2021MTNR}
Z.~Wang, S.~Wende-von Berg, and M.~Braun.
\newblock Fast parallel {Newton–Raphson} power flow solver for large number
  of system calculations with {CPU} and {GPU}.
\newblock {\em Sustainable Energy, Grids and Networks}, 27:100483, Sep 2021.

\end{thebibliography}

\vfill
\begin{flushright}
{\footnotesize
  \framebox{\parbox{0.5\textwidth}{
The submitted manuscript has been created by UChicago Argonne, LLC,
Operator of Argonne National Laboratory (``Argonne"). Argonne, a
U.S. Department of Energy Office of Science laboratory, is operated
under Contract No. DE-AC02-06CH11357. The U.S. Government retains for
itself, and others acting on its behalf, a paid-up nonexclusive,
irrevocable worldwide license in said article to reproduce, prepare
derivative works, distribute copies to the public, and perform
publicly and display publicly, by or on behalf of the Government.
The Department of
Energy will provide public access to these results of federally sponsored research in accordance
with the DOE Public Access Plan. http://energy.gov/downloads/doe-public-access-plan. }}
\normalsize
}
\end{flushright}

\end{document}